\documentclass[aps,prl,twocolumn,showpacs,superscriptaddress,floatfix]{revtex4-1}

\usepackage{graphicx}
\usepackage{dcolumn}
\usepackage{bm}
\usepackage{amssymb}
\usepackage{microtype}
\usepackage{xfrac}
\usepackage{array}

\newcommand{\ket}[1]{\ensuremath{|#1\rangle}}

\begin{document}

\title{XY Antiferromagnetic Ground State in the Effective $S=\sfrac{1}{2}$ Pyrochlore Yb$_2$Ge$_2$O$_7$}

\author{A.~M.~Hallas}
\affiliation{Department of Physics and Astronomy, McMaster University, Hamilton, ON, L8S 4M1, Canada}

\author{J.~Gaudet}
\affiliation{Department of Physics and Astronomy, McMaster University, Hamilton, ON, L8S 4M1, Canada}

\author{M.~N.~Wilson}
\affiliation{Department of Physics and Astronomy, McMaster University, Hamilton, ON, L8S 4M1, Canada}

\author{T.~J.~Munsie}
\affiliation{Department of Physics and Astronomy, McMaster University, Hamilton, ON, L8S 4M1, Canada}

\author{A.~A.~Aczel}
\affiliation{Quantum Condensed Matter Division, Oak Ridge National Laboratory, Oak Ridge, Tennessee 37831, USA}

\author{M.~B.~Stone}
\affiliation{Quantum Condensed Matter Division, Oak Ridge National Laboratory, Oak Ridge, Tennessee 37831, USA}

\author{R.~S.~Freitas}
\affiliation{Instituto de F\'{i}sica, Universidade de S\~{a}o Paulo, S\~{a}o Paulo 05315-970, SP, Brazil}

\author{A.~M.~Arevalo-Lopez}
\affiliation{Centre for Science at Extreme Conditions and School of Chemistry, University of Edinburgh, Peter Guthrie Tait Road, King's Buildings, Edinburgh EH9 3FD, United Kingdom}

\author{J.~P.~Attfield}
\affiliation{Centre for Science at Extreme Conditions and School of Chemistry, University of Edinburgh, Peter Guthrie Tait Road, King's Buildings, Edinburgh EH9 3FD, United Kingdom}

\author{M.~Tachibana}
\affiliation{National Institute for Materials Science, 1-1 Namiki, Tsukuba 305-0044, Ibaraki, Japan}

\author{C.~R.~Wiebe}
\affiliation{Department of Chemistry, University of Winnipeg, Winnipeg, MB, R3B 2E9 Canada}
\affiliation{Canadian Institute for Advanced Research, 180 Dundas St. W., Toronto, ON, M5G 1Z7, Canada}

\author{G.~M.~Luke}
\affiliation{Department of Physics and Astronomy, McMaster University, Hamilton, ON, L8S 4M1, Canada}
\affiliation{Canadian Institute for Advanced Research, 180 Dundas St. W., Toronto, ON, M5G 1Z7, Canada}

\author{B.~D.~Gaulin}
\affiliation{Department of Physics and Astronomy, McMaster University, Hamilton, ON, L8S 4M1, Canada}
\affiliation{Canadian Institute for Advanced Research, 180 Dundas St. W., Toronto, ON, M5G 1Z7, Canada}
\affiliation{Brockhouse Institute for Materials Research, Hamilton, ON L8S 4M1 Canada}

\date{\today}

\begin{abstract} 
We report neutron scattering and muon spin relaxation measurements ($\mu$SR) on the pyrochlore antiferromagnet Yb$_2$Ge$_2$O$_7$. Inelastic neutron scattering was used to probe the transitions between crystal electric field levels, allowing us to determine the eigenvalues and eigenvectors appropriate to the $J=\sfrac{7}{2}$ Yb$^{3+}$ ion in this environment. The crystal electric field ground state doublet in Yb$_2$Ge$_2$O$_7$ corresponds primarily to $m_J=\pm \sfrac{1}{2}$ with local XY anisotropy, consistent with an $S_{\text{eff}}=\sfrac{1}{2}$ description for the Yb moments. $\mu$SR measurements reveal the presence of an ordering transition at $T_N=0.57$~K with persistent weak dynamics in the ordered state. Finally, we present neutron diffraction measurements that reveal a clear phase transition to the $\mathbf{k}=(000)$ $\Gamma_5$ ground state with an ordered magnetic moment of 0.3(1) $\mu_B$ per Yb ion. We compare and contrast this phenomenology with the low temperature behavior of Yb$_2$Ti$_2$O$_7$ and Er$_2$Ti$_2$O$_7$, the prototypical $S_{\text{eff}}=\sfrac{1}{2}$ XY pyrochlore magnets. 
\end{abstract}


\maketitle

\section{Introduction}

Magnetic frustration arises for systems in which the lattice geometry precludes the simultaneous satisfaction of all pairwise magnetic interactions. Cubic pyrochlore oxides, with the composition $A_2B_2$O$_7$, present exemplary three dimensional realizations of lattices that can be subject to strong geometric magnetic frustration when either the $A$ or $B$ site is occupied by a magnetic cation \cite{RevModPhys.82.53}. The sublattices produced by the $A$ and $B$ site cations form two interpenetrating networks of corner-sharing tetrahedra. The topicality of the pyrochlore lattice for the study of magnetic frustration is, in part, due to the ease with which numerous magnetic and non-magnetic cations can be substituted onto the $A$ and $B$ sites of the lattice \cite{Subramanian198355}. As a result, a plethora of magnetic pyrochlores have been investigated, revealing a diverse array of exotic magnetic ground states. 

Rare-earth titanates and stannates of the form $R_2B_2$O$_7$ where $R$ is a rare-earth ion, and non-magnetic $B$ is either Ti$^{4+}$ or Sn$^{4+}$ have been of great experimental interest. Both of these families can be synthesized using a wide range of rare-earth ions. However, while it is straightforward to grow large single crystals of the titanate $R_2$Ti$_2$O$_7$ series, the stannate $R_2$Sn$_2$O$_7$ series exists only in powder form at present. More recently, the rare-earth germanate family, $R_2$Ge$_2$O$_7$, has presented a new avenue to investigate the physics of magnetic pyrochlores. The germanate family is relatively unexplored as they can only be grown under high pressures, and have thus far only been obtained as small polycrystalline samples \cite{:/content/aip/journal/aplmater/3/4/10.1063/1.4916020}. The germanate pyrochlores, due to the small ionic radius of Ge$^{4+}$, have contracted lattice parameters with respect to their titanium and tin analogs and thus, have so far been studied in the context of chemical pressure \cite{:/content/aip/journal/aplmater/3/4/10.1063/1.4916020,PhysRevLett.113.267205,PhysRevB.89.064401}.

The diversity of magnetic ground states observed across the $R_2B_2$O$_7$ series, with $B=$~Ge, Ti, or Sn, can be primarily attributed to two sources. Firstly, the moment size and anisotropy differ significantly, depending upon which rare-earth sits at the $A$-site. These single ion properties are determined by the crystal field splitting of the (2$J$+1) multiplet arising from the partially-filled 4$f$ shell at the $R^{3+}$ site. Secondly, the relative strength and nature of the magnetic interactions that exist between the $R^{3+}$ moments can vary greatly. Furthermore, due to strong spin-orbit coupling in the 4$f$ series, the exchange interactions between the $R^{3+}$ moments are anisotropic, and the form of these interactions is determined by the point group symmetry at the $R^{3+}$ site \cite{PhysRevB.78.094418}. In simple terms, these combinations can generate ferromagnetically or antiferromagnetically coupled Ising, XY, or Heisenberg spins decorating a network of corner-sharing tetrahedra, and the diversity of ground states that these combinations imply. 

While a range of magnetic ground states exist in the rare-earth pyrochlores, we limit ourselves henceforth to discussion of those rare-earth pyrochlores with XY anisotropy, specifically the Yb$^{3+}$ and Er$^{3+}$ pyrochlores. The crystal electric field states for  Yb$_2$Ti$_2$O$_7$ are well understood \cite{YTO8,YTO9,YTO10}, while those corresponding to Er$_2$Ti$_2$O$_7$ are less well-determined \cite{ETO2,ETO16}. However, it is clear that both the Yb$^{3+}$ and Er$^{3+}$ ions in $R_2$Ti$_2$O$_7$ give rise to XY-like magnetic anisotropy. This XY anisotropy implies that the eigenfunctions describing the ground state doublet have large contributions from $m_J=\pm \sfrac{1}{2}$. Provided that the ground state doublet is well-separated from the first excited crystal field level, this results in an $S_{\text{eff}}=\sfrac{1}{2}$ quantum description for the magnetic degrees of freedom. Similarities in crystal structure and associated crystal field effects suggest that the same should be true for all of Yb$_2B_2$O$_7$ and Er$_2B_2$O$_7$ with $B=$~Ge, Ti, and Sn. 

Yb$_2$Ti$_2$O$_7$ and Yb$_2$Sn$_2$O$_7$ both possess Curie-Weiss constants that are ferromagnetic and weak \cite{ETO1,YTO9,YSO1}. These two materials have also both been found to order into a canted ferromagnetic state at low temperatures \cite{Chang,Yasui,Jonathan_YTO,YSO1}. However, there are exotic characteristics to such states, at least in the case of Yb$_2$Ti$_2$O$_7$, for which single crystal studies are required. For example in the ``ordered'' state of Yb$_2$Ti$_2$O$_7$ there is persistent anisotropic diffuse scattering \cite{YTO3,PhysRevLett.106.187202} and no evidence of well-defined spin wave excitations in zero magnetic field \cite{YTO4,YTO5}. Furthermore, single crystal inelastic neutron scattering measurements of Yb$_2$Ti$_2$O$_7$ in its field-induced polarized state have been used to estimate its microscopic spin Hamiltonian \cite{YTO1,YTO2}. Interestingly, the heat capacity anomalies in Yb$_2$Ti$_2$O$_7$ are known to be sample dependent with sensitivity to the presence of weak quenched disorder at the 1\% level \cite{YTO6,YTO7}. This sample dependence also extends to the ground state properties, as observed with both $\mu$SR and neutron scattering \cite{PhysRevB.88.134428,PhysRevB.89.184416,YTO4,YTO5,Chang,Yasui}.

Er$_2$Ti$_2$O$_7$ is an interesting contrast to the ytterbium systems. Er$_2$Ti$_2$O$_7$ is known to possess a relatively large, antiferromagnetic Curie-Weiss constant \cite{ETO1,ETO2} and undergoes a continuous phase transition to a non-coplanar $\psi_2$ antiferromagnetic ordered state at $T_N=$1.2~K \cite{ETO8,ETO3,ETO4}. However, in contrast to Yb$_2$Ti$_2$O$_7$, there are well-defined conventional spin wave excitations in Er$_2$Ti$_2$O$_7$ below $T_N$ \cite{ETO5}. Its microscopic spin Hamiltonian has also been estimated from inelastic neutron scattering \cite{ETO6}, and the selection of the $\psi_2$ ground state is argued to arise due to an order-by-quantum disorder mechanism \cite{ETO6,ETO7,ETO8,ETO9,ETO14,ETO15}. The corresponding order-by-disorder spin wave gap has been measured with inelastic neutron scattering \cite{ETO10}. An alternate energetic argument for ground state selection of the $\psi_2$ state has also recently been made \cite{ETO12,ETO13}. In striking contrast to Yb$_2$Ti$_2$O$_7$, the $\psi_2$ ground state in Er$_2$Ti$_2$O$_7$ is not obviously sensitive to quenched disorder and does not display sample dependence. It has even been shown to accommodate magnetic dilution consistent with 3D percolation theory \cite{ETO11}.

In this paper, we turn our attention to a member of the germanate pyrochlore family, Yb$_2$Ge$_2$O$_7$, wherein Ge$^{4+}$ on the $B$-site is non-magnetic and the magnetism is carried by Yb$^{3+}$ on the $A$-site. We first present our inelastic neutron scattering measurements, which establish the eigenvalues and eigenfunctions for the crystal field levels appropriate to Yb$_2$Ge$_2$O$_7$. This firmly establishes the XY nature of the Yb$^{3+}$ moments in their $g$-tensor anisotropy. We next show $\mu$SR measurements that establish a phase transition at $T_N=0.57$~K to a conventional long-range ordered state with weak dynamics. Finally we present elastic neutron scattering measurements which reveal the ordered state in Yb$_2$Ge$_2$O$_7$ to be the $\mathbf{k}=(000)$ $\Gamma_5$ antiferromagnetic structure with an ordered moment of 0.3(1)~$\mu_{\text{B}}$. As both Yb$_2$Ge$_2$O$_7$ and Er$_2$Ti$_2$O$_7$ are antiferromagnetically-coupled systems with ordered states in the $\Gamma_5$ manifold, $S_{\text{eff}}=\sfrac{1}{2}$ degrees of freedom, and XY anisotropy, we compare and contrast these two pyrochlores.

\section{Experimental Methods}

The cubic pyrochlore phase of Yb$_2$Ge$_2$O$_7$ cannot be stabilized at ambient pressure using conventional solid state synthesis. Thus, powder samples of Yb$_2$Ge$_2$O$_7$ were synthesized using a belt-type high pressure apparatus. Stoichiometric quantities of Yb$_2$O$_3$ and GeO$_2$, pre-reacted into the tetragonal pyrogermanate phase, were sealed in gold capsules and reacted at 1300$^{\circ}$C under 6~GPa of pressure. The resulting product was thoroughly ground and powder x-ray diffraction was used to confirm the $Fd\bar{3}m$ pyrochlore structure for each 400~mg batch. Our Rietveld refinements of the x-ray patterns gave a lattice parameter of 9.8284(2)~\AA, in agreement with previous reports \cite{YGOStructure,PhysRevB.89.064401}. While scaling up materials synthesized under high pressure is cumbersome, such samples do present some inherent advantages. Firstly, high pressure synthesis gives a high degree of control over the stoichiometry \cite{PhysRevLett.113.267205}. Furthermore, the large ionic radii difference between Yb$^{3+}$ and Ge$^{4+}$, which necessitates high-pressure synthesis, also significantly reduces the probability of site-mixing \cite{:/content/aip/journal/aplmater/3/4/10.1063/1.4916020}. This is particularly attractive in light of the sensitivity that the magnetism in some pyrochlores has shown to subtle variations in stoichiometry and so-called ``stuffing'' \cite{YTO6, PhysRevB.87.060408, arXiv:1509.04583}.

Muon spin relaxation ($\mu$SR) measurements on Yb$_2$Ge$_2$O$_7$ were carried out at the TRIUMF Laboratory. A 300~mg pressed pellet of Yb$_2$Ge$_2$O$_7$, mixed with 20\% silver powder to improve thermalization, was attached to a silver coated cold finger with Apiezon N-grease. Measurements between 25~mK and 2~K were carried out in a dilution refrigerator, both with zero external field and in fields up to 0.5~T. In $\mu$SR measurements, 100\% spin polarized muons are implanted one at a time in a sample, where the muon spins evolve in the local magnetic environment. As the muons decay, with an average lifetime of 2.2~$\mu$s, they emit a positron preferentially in the direction of the muon spin. Two opposing sets of detectors, in the forward and backward direction from the sample relative to the muon beam, detect the emitted positrons. The asymmetry spectra, which is directly proportional to the muon polarization, is described by $A(t) = [F(t)-B(t)]/[F(t)+B(t)]$, where F(t) and B(t) are the number of positrons detected in the forward and backward direction respectively, scaled by their counting efficiencies. Fits to the $\mu$SR data were performed using musrfit \cite{Suter201269}.   

The inelastic neutron scattering measurements on Yb$_2$Ge$_2$O$_7$ were performed on the SEQUOIA spectrometer at the Spallation Neutron Source at Oak Ridge National Laboratory. A 2.7 gram powder sample of Yb$_2$Ge$_2$O$_7$ was sealed in an aluminum sample can under a helium atmosphere. Using a standard orange ILL cryostat, measurements were performed at 2~K with an incident energy of 150~meV, giving an elastic energy resolution of $\pm$2.8~meV. The energy resolution improves at higher energies and is approximately 1.4~meV for energy transfers of 80~meV and and 1~meV for energy transfers of 120~meV. This configuration was also used for measurements on an identical empty can, which serves as a background.

Magnetic neutron diffraction measurements on Yb$_2$Ge$_2$O$_7$ were carried out with the fixed incident energy triple axis spectrometer HB1A at the High Flux Isotope Reactor at Oak Ridge National Laboratory. The same 2.7 gram sample of Yb$_2$Ge$_2$O$_7$ was mounted in an oxygen free copper sample can under a helium atmosphere. The incident neutron beam has a wavelength of 2.37~\AA, which is selected by a double pyrolitic graphite monochromator. Energy analysis of the scattered beam employs a pyrolitic graphite analyzer crystal, giving an elastic energy resolution of approximately 1~meV. Elastic diffraction measurements were carried out using both a $^3$He insert and a dilution insert, with base temperatures of 300~mK and 50~mK, respectively, and a maximum temperature of 10~K. Representational analysis of the diffracted intensities were performed using SARAh Refine \cite{Wills2000680} and FullProf \cite{RodriguezCarvajal199355}. 


\section{Results and Discussion}

\subsection{I. Determination of the Crystal Electric Field Eigenvalues and Eigenfunctions via Inelastic Neutron Scattering}

\begin{figure}[tbp]
\linespread{1}
\par
\includegraphics[width=3.3in]{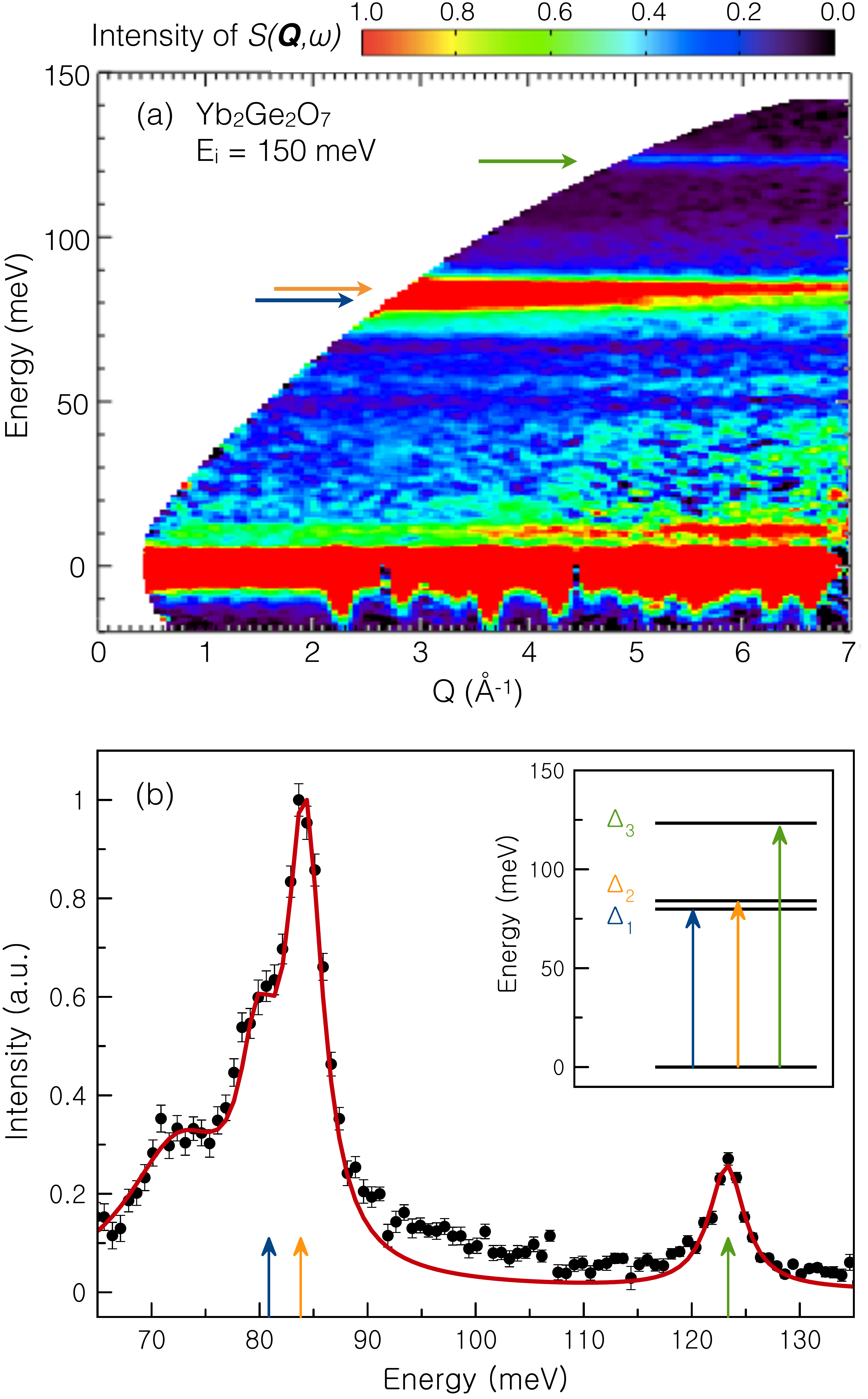}
\par
\caption{(a) Inelastic neutron scattering spectrum of Yb$_2$Ge$_2$O$_7$ measured at 2~K with a neutron beam of incident energy 150~meV. A background spectrum, measured on an empty can, has been subtracted from the data. Crystal field excitations at 80.7~meV, 84.2~meV, and 123.3~meV are indicated by the blue, yellow, and green arrows, respectively. (b) An integrated cut of the same data over the range $Q=$~[5.4, 6.0] \AA$^{-1}$. The fit to the data is achieved using a Hamiltonian, given by Eqn.~1, which approximates the Coulomb potential generated by the crystal electric field due to the neighboring oxygen atoms.}
\label{YGOCEF}
\end{figure}

\begin{figure}[tbp]
\linespread{1}
\par
\includegraphics[width=3.3in]{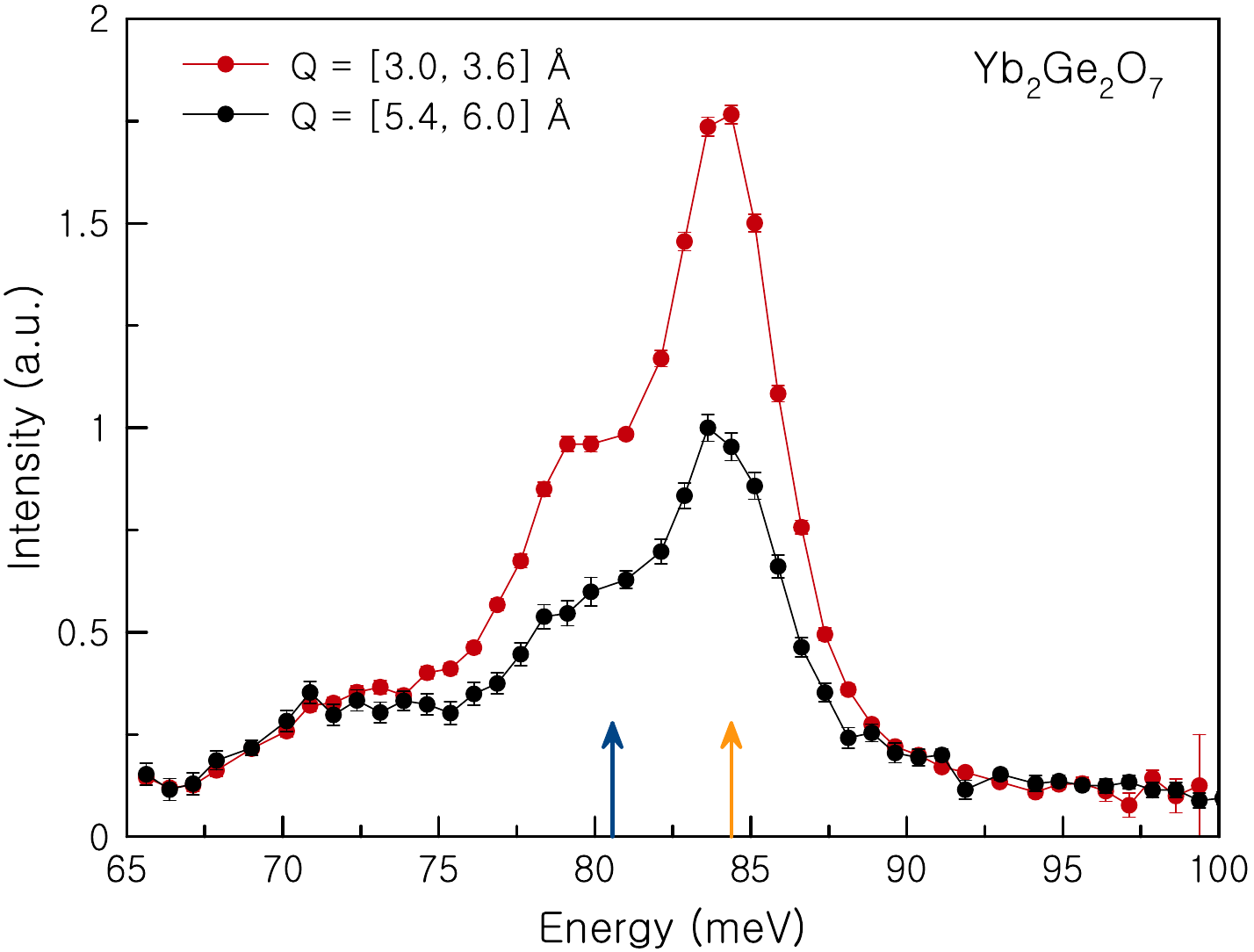}
\par
\caption{The $Q$ dependence of the first two crystal electric field levels in Yb$_2$Ge$_2$O$_7$, indicated by the blue and yellow arrows respectively. The decrease in intensity as a function of $Q$ is consistent with the magnetic form factor for Yb$^{3+}$.}
\label{YGOCEFQ}
\end{figure}

Figure \ref{YGOCEF}(a) shows the inelastic neutron scattering spectrum for Yb$_2$Ge$_2$O$_7$ collected at 2~K for energy transfers up to 150~meV. The excitations corresponding to transitions between crystal electric field (CEF) levels can be assigned based on two criteria: (i) They should be dispersionless, \emph{ie.} without $Q$-dependence and (ii) Their intensity should be maximal at the lowest $Q$-values and should fall off according to the magnetic form factor of Yb$^{3+}$. Following these criteria, three crystal field excitations can be identified in Figure~\ref{YGOCEF}(a) at 80.7~meV, 84.2~meV, and 123.3~meV. 
The $Q$ dependence for the first two transitions is shown in Figure~\ref{YGOCEFQ} and is consistent with the Yb$^{3+}$ magnetic form factor. The valence shell of Yb$^{3+}$ contains 13 $f$ electrons which, following Hund's rules, gives a spin orbit ground state with total angular momentum $J=\sfrac{7}{2}$ and a $2J+1=8$ fold degeneracy. The local oxygen environment surrounding each Yb$^{3+}$ cation produces a crystal electric field that lifts the ground state degeneracy. However, from its odd electron count, it follows that Yb$^{3+}$ is subject to Kramer's theorem and, consequently, the crystal electric field can produce at most four doublets. Thus, the three crystal field doublets observed in Figure~\ref{YGOCEF}(a) and the ground state doublet account for the full manifold of the Yb$^{3+}$ crystal electric field transitions.

The eight oxygens that surround each Yb$^{3+}$ cation form a cube that is distorted along one of its body diagonals, where this direction forms the local [111] axis. Defining the $\hat{z}$ axis along this [111] local axis, the local environment has a 3-fold symmetry, as well as an inversion symmetry, giving a point group symmetry $D_{3d}$. Following the Stevens' operator formalism, we use the following Hamiltonian to approximate the Coulomb potential generated by the crystal electric field due to the neighboring oxygen atoms \cite{Prather,Stevens,Hutchings}:
\begin{eqnarray}
\mathcal{H}_{\text{CEF}}= B^0_2\hat{O}^0_2 + B^0_4\hat{O}^0_4 + B^3_4\hat{O}^3_4 +
\nonumber
\\
B^0_6\hat{O}^0_6 + B^3_6\hat{O}^3_6 + B^6_6\hat{O}^6_6
\label{eq: HCEF}
\end{eqnarray} 
where the CEF parameters, $B^m_n$, have been fit using our inelastic neutron scattering data of Figure~\ref{YGOCEF}(a). To do so, the scattered intensity is computed in the same way as in Ref.~\cite{YTO8} using the CEF parameters of Yb$_2$Ti$_2$O$_7$ as an initial guess. This calculation is compared by a least squares refinement to the experimental result (in this case, an integrated cut over $Q=$~[5.4, 6.0] \AA$^{-1}$, as shown in Figure~\ref{YGOCEF}(b)). The CEF parameters are varied until good agreement is obtained with the experimental data. The resulting fit is shown in Figure~\ref{YGOCEF}(b) where a Lorentzian function has been added at 73~meV to phenomenologically capture the scattered intensity produced from a non-magnetic contribution, likely phonon or multiple phonon scattering. The non-magnetic origin of this feature can be deduced from the lack of $Q$ dependence, as shown in Figure~\ref{YGOCEFQ}, as well as the temperature dependence, as was shown for Yb$_2$Ti$_2$O$_7$ \cite{YTO8}.

The resulting CEF parameters are: $B^0_2=1.08$~meV, $B^0_4=-6.32\cdot10^{-2}$~meV, $B^3_4=3.02\cdot10^{-1}$~meV, $B^0_6=9.25\cdot10^{-4}$~meV, $B^3_6=4.66\cdot10^{-2}$~meV, and $B^6_6=3.10\cdot10^{-3}$~meV. The corresponding eigenfunctions and eigenvalues obtained with the above CEF parameters are given in Table~\ref{CEFeig}. As is the case for Yb$_2$Ti$_2$O$_7$, the CEF ground state is primarily comprised of $m_J$~=~$\pm \sfrac{1}{2}$. As well, the first, second, and third excited states of Yb$_2$Ge$_2$O$_7$ are predominantly made up of $m_J$~=~$\pm \sfrac{7}{2}$, $m_J$~=~$\pm \sfrac{3}{2}$ and $m_J$~=~$\pm \sfrac{5}{2}$, respectively. In fact, the crystal electric field scheme of Yb$_2B_2$O$_7$ appears to be relatively unperturbed by substitution of the non-magnetic $B$-site from Ti$^{4+}$ to Ge$^{4+}$. Each of the excited crystal field states in Yb$_2$Ge$_2$O$_7$ is shifted upwards in energy approximately 5\% from the corresponding level in Yb$_2$Ti$_2$O$_7$. This can be understood intuitively in terms of the reduction in the lattice parameter going from Ti$^{4+}$ to Ge$^{4+}$. The reduced lattice parameter leads to a contraction of the oxygen atoms about the Yb$^{3+}$ cations and, as a result, Yb$^{3+}$ feels a larger crystal electric field, leading to a stronger splitting of the levels.

The anisotropic $g$-tensors for Yb$_2$Ge$_2$O$_7$, obtained from the fit shown in Figure~\ref{YGOCEF}(b), are $g_{\perp}$~=~3.5(2) and $g_z$~=~2.1(1), where \textit{z} corresponds to the local [111] axis. The uncertainty on the $g$-tensors have been obtained by comparing the best fit results with and without the phonon contribution around 73~meV and with and without a sloping background taken along the full energy range. The resulting $g$-tensor anisotropy corresponds to XY anisotropy and can be quantified by taking their ratio, giving $g_{\perp}/g_z = 1.7(2)$. For comparison, the value obtained for the same ratio with Yb$_2$Ti$_2$O$_7$ is 1.9(2) \cite{YTO8}, indicating that the XY anisotropy may be slightly stronger in Yb$_2$Ti$_2$O$_7$ as compared to Yb$_2$Ge$_2$O$_7$. 

\begin{table} 
\begin{tabular}{|c||c|c|c|c|c|c|c|c|}
\toprule
E (meV) & $\ket{\text{-}\sfrac{7}{2}}$ & $\ket{\text{-}\sfrac{5}{2}}$ & $\ket{\text{-}\sfrac{3}{2}}$ & $\ket{\text{-}\sfrac{1}{2}}$ & $\ket{\sfrac{1}{2}}$ & $\ket{\sfrac{3}{2}}$ & $\ket{\sfrac{5}{2}}$ & $\ket{\sfrac{7}{2}}$ \\
\colrule
0 & 0 & 0.13 & 0 & 0 & -0.91 & 0 & 0 & 0.40 \\
0 & -0.40 & 0 & 0 & -0.91 & 0 & 0 & -0.13 & 0 \\
80.7 & 0.90 & 0 & 0 & 0.36 & 0 & 0 & -0.24 & 0 \\
80.7 & 0 & -0.24 & 0 & 0 & 0.36 & 0 & 0 & 0.90 \\
84.2 & 0 & 0 & -1 & 0 & 0 & 0 & 0 & 0 \\
84.2 & 0 & 0 & 0 & 0 & 0 & -1 & 0 & 0 \\
123.3 & 0.05 & -0.93 & 0 & -0.05 & -0.21 & 0 & 0.23 & -0.17 \\
123.3 & 0.17 & 0.23 & 0 & -0.21 & 0.05 & 0 & 0.93 & 0.05 \\
\botrule
\end{tabular}
\caption{The crystal electric field (CEF) eigenvalues and eigenvectors for Yb$^{3+}$ at the $A$-site of Yb$_2$Ge$_2$O$_7$. The first column displays the CEF eigenvalues of the system, while the corresponding eigenvectors are given in each row in terms of the $m_J$ basis.}
\label{CEFeig}
\end{table}


\subsection{II. Static Magnetism Revealed by Muon Spin Relaxation}

\begin{figure}[htbp]
\linespread{1}
\par
\includegraphics[width=3.3in]{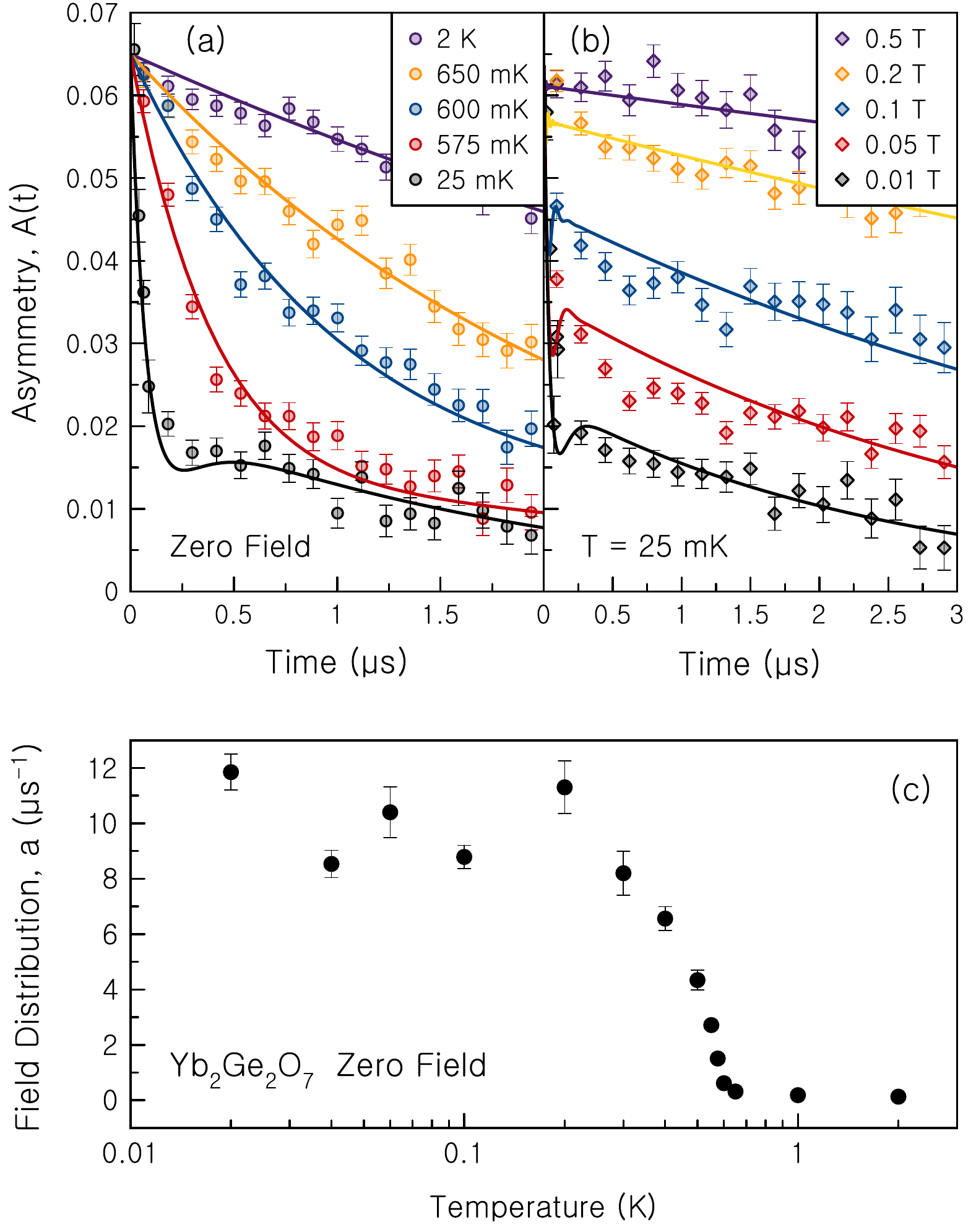}
\par
\caption{The results from muon spin relaxation measurements performed on Yb$_2$Ge$_2$O$_7$ between 25~mK and 2~K (a) Several representative asymmetry spectra for Yb$_2$Ge$_2$O$_7$ in which the background asymmetry has been subtracted. The fits to the data are indicated by the solid lines. The asymmetry is fit by a dynamical Lorentzian Kubo-Toyabe function. (b) The longitudinal field decoupling at at 25~mK in fields between 0.01~T and 0.5~T. The fits, as indicated by the solid lines, are given by a dynamical Lorentzian Kubo-Toyabe function with the magnitude of the external field imposed as a constraint. (c) The fitted values for the internal field distribution, $a$, an adjustable parameter in the dynamical Lorentzian Kubo-Toyabe function. The field distribution forms an order parameter which plateaus below 200~mK, and shows the onset of magnetic order at $T_N=0.57$~K.}
\label{YGOmuSR}
\end{figure}

We employed muon spin relaxation ($\mu$SR) measurements to further characterize the low temperature magnetism in Yb$_2$Ge$_2$O$_7$. 
Figure~\ref{YGOmuSR}(a) shows some representative asymmetry spectra for Yb$_2$Ge$_2$O$_7$ in zero external field between 0 and 2~$\mu$s. A background asymmetry contribution was fit to a slowly relaxing temperature independent exponential and then subtracted from the presented data. This background asymmetry contribution comes from muons that land outside the sample, either in the admixed silver, the silver sample holder, or the cryostat. 

At sufficiently high temperatures, when a system is in its paramagnetic regime, there will only be a small relaxation due to nuclear dipole moments; at such temperatures, the spins in the sample are rapidly fluctuating and the dynamics are faster than the muon time window. In Yb$_2$Ge$_2$O$_7$ we see that at 2~K the asymmetry is only weakly relaxing, indicating that the sample is within its paramagnetic regime. At 1~K, there is a slight increase in the relaxation, indicating that electronic spin correlations are beginning to develop. Upon cooling towards the N\'eel temperature, the relaxation further increases due to slowing fluctuations as the electronic correlations grow stronger (Figure~\ref{YGOmuSR}(a)). As Yb$_2$Ge$_2$O$_7$ is cooled below the N\'eel temperature, the asymmetry takes on a two component form, with a sharp drop in the early time asymmetry followed by a slow relaxation at longer times.

The data at all temperatures are well described in terms of a dynamical Lorentzian Kubo-Toyabe function \cite{Tomo}. This function, which is appropriate for a system with a Lorentzian distribution of internal fields, $a$, appropriately captures the physics in a system with slow or fast dynamics. The temperature evolution of $a$, as shown in Figure~\ref{YGOmuSR}(c), provides a clear order parameter corresponding well with the observed N\'eel temperature from ac susceptibility and heat capacity, $T_N=0.57$~K \cite{PhysRevB.89.064401}. The internal field distribution plateaus below 200~mK at approximately 12~$\mu$s$^{-1}$, which corresponds to 0.1~T.   

The asymmetry spectra for Yb$_2$Ge$_2$O$_7$ differs from the canonical spectra for a system with static magnetic order in two key aspects. Firstly, the long time component is not fully time-independent but instead has a weak exponential relaxation. This indicates that below the N\'eel temperature, Yb$_2$Ge$_2$O$_7$ remains weakly dynamic. A persistent relaxing signal, while not fully understood, is a common feature of magnetically frustrated systems \cite{PhysRevB.91.104427}. Secondly, no long-lived precessing signal could be resolved in the asymmetry spectra. However, a lack of oscillations does not preclude static magnetic order. In fact, the absence of oscillations is frequently observed for pyrochlores with long range magnetic order, such as Er$_2$Ti$_2$O$_7$ \cite{0953-8984-17-6-015} and Tb$_2$Sn$_2$O$_7$ \cite{PhysRevLett.96.127202,PhysRevLett.97.117203}. A lack of oscillations can be attributed to an inhomogeneous internal field distribution, which can, in part, be explained by having multiple muon stopping sites. Similarly, the minimum in Kubo Toyabe function can be ``wiped out'' by multiple field distributions as would be expected for multiple muon stopping sites \cite{PhysRevB.56.2352}. As the pyrochlore structure contains two inequivalent oxygen sites, and positively charged muons stop at the most electronegative positions, at least two inequivalent stopping sites can be expected in Yb$_2$Ge$_2$O$_7$. 

We also performed $\mu$SR measurements on Yb$_2$Ge$_2$O$_7$ with an externally applied magnetic field. In our measurements, the external field is applied parallel to the initial muon polarization direction, \emph{ie.} longitudinal geometry. In the case of static (or quasi-static) magnetism, the external field can be increased until it overwhelms the static internal fields produced by the sample. When this happens, the muon spins will respond more strongly to the external field and become effectively ``decoupled'' from the sample, resulting in a reduced relaxation rate. For Yb$_2$Ge$_2$O$_7$, at 25~mK in fields between 0.01~T and 0.5~T, the asymmetry decouples in the expected manner for a dynamic Lorentzian Kubo-Toyabe, with the magnitude of the external field imposed as a constraint (Figure~\ref{YGOmuSR}(b)). While applying a longitudinal field effectively decouples the majority of the relaxation, as in the zero field case, there remains a weak long time relaxation, so-called persistent spin dynamics. Thus, both the zero field and longitudinal field measurements on Yb$_2$Ge$_2$O$_7$ are consistent with quasi-static magnetic order on the muon timescale.


\subsection{III. Measurement of the Magnetic Structure by Elastic Neutron Diffraction}

\begin{figure}[htbp]
\linespread{1}
\par
\includegraphics[width=3.3in]{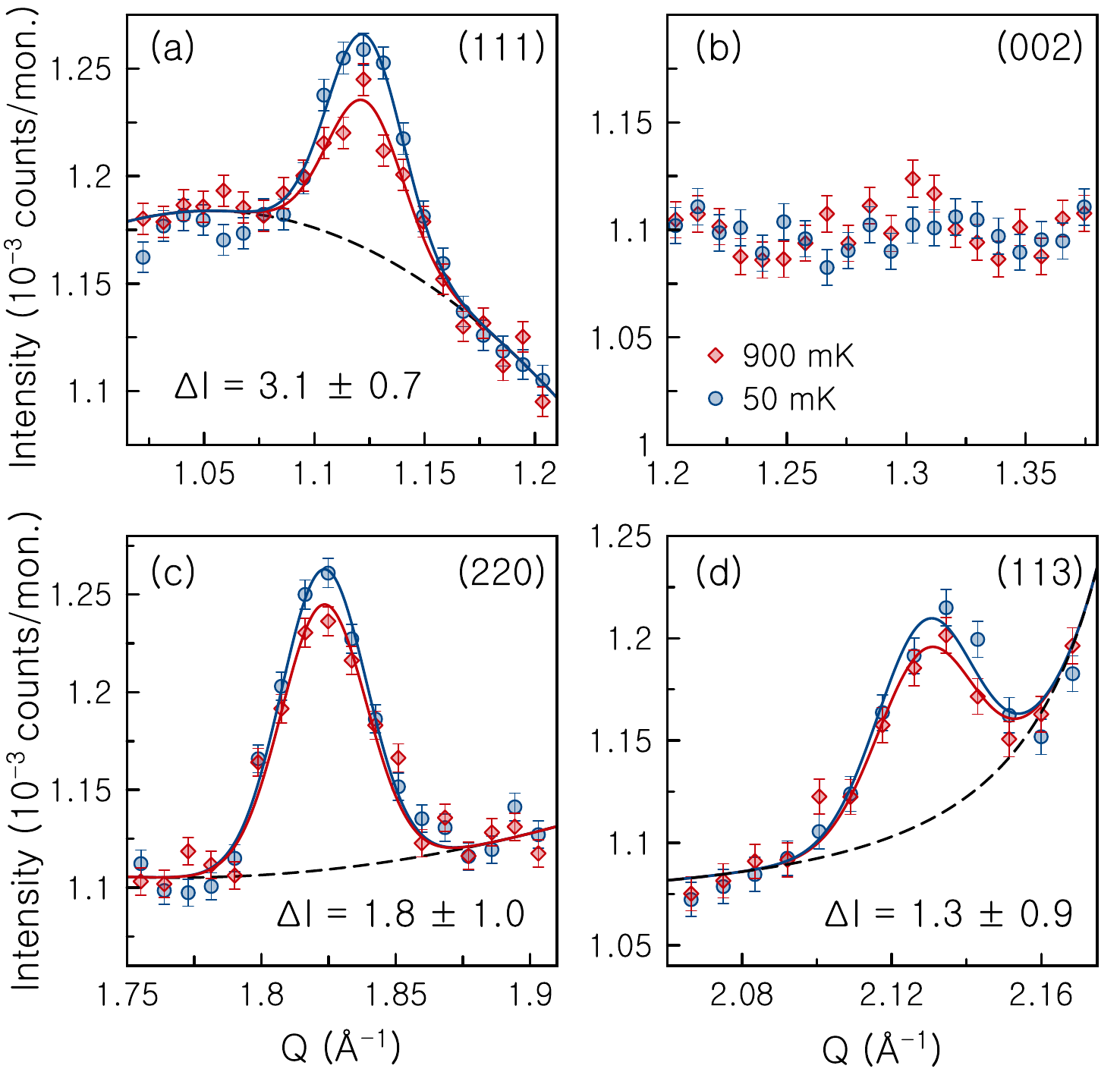}
\par
\caption{Neutron diffraction of Yb$_2$Ge$_2$O$_7$ over regions of $Q$-space corresponding to the (a) (111) (b) (002) (c) (220) and (d) (113) nuclear-allowed Bragg positions. Data collected below $T_N$ (50~mK) is shown in blue and above $T_N$ (900~mK) in red. The Bragg peaks are fit by Lorentzian functions in which the peak position and the background (indicated by the dashed black line) are jointly refined. The peak widths were fixed by fitting the (222) nuclear reflection. Thus, for each Bragg reflection, the only independent adjustable parameter is the peak area. The intensity gain on cooling into the N\'eel state is indicated on each panel, denoted as $\Delta I$, except in the case of (002) where no reflection is observed. The relative intensities of the observed magnetic Bragg reflections in Yb$_2$Ge$_2$O$_7$ correspond to a $\Gamma_5$ ordered state.}
\label{YGOBragg}
\end{figure}

Magnetic neutron diffraction was employed to determine the magnetic ground state of Yb$_2$Ge$_2$O$_7$. Our initial measurements surveyed a broad region of $Q$-space between 0.5~\AA$^{-1}$~and 2.5~\AA$^{-1}$. Comparison of data sets collected at 50~mK and 900~mK revealed the formation of magnetic Bragg peaks on cooling into the ordered phase. All magnetic Bragg peaks were observed to form on allowed positions for nuclear reflections in the pyrochlore lattice. Figure~\ref{YGOBragg} shows four of the measured Bragg positions at 50~mK and 900~mK. In each case, the peak has been fit to a Lorentzian peak shape function where the only independent adjustable parameter between 50~mK and 900~mK is the peak area. The peak centers and the background (denoted by the dashed line) were jointly refined. The peak widths were fixed according to the width of the (222) nuclear peak, which is the largest nuclear reflection. 

Below the N\'eel transition in Yb$_2$Ge$_2$O$_7$, the largest intensity gain is observed on the (111) position (Figure~\ref{YGOBragg}(a)). The intensity gain from 900~mK to 50~mK, in arbitrary units of counts normalized by monitor, is 3.1 $\pm$ 0.7. The (002) position, which does not contain a nuclear reflection, is also devoid of a magnetic reflection (Figure~\ref{YGOBragg}(b)). The next magnetic reflections occur at (220) and (113), which have intensity gains of 1.8 $\pm$ 1.0 and 1.3 $\pm$ 0.9, respectively (Figure~\ref{YGOBragg}(c,d)). The (113) Bragg peak, which is centered at 2.13~\AA$^{-1}$ has a large sloping background because it is immediately adjacent to the large (222) nuclear reflection, centered at 2.23~\AA$^{-1}$. The (222) and (004) positions were also carefully measured and showed no intensity gain below the N\'eel temperature. A summary of the experimental intensities, given as a fraction of the intensity of (111), are shown in Table~\ref{Refinement}.

\begin{table}[tbp]
\begin{tabular}{|l||c|c|c|c|c|c|}
\toprule
 & (111) & (002) & (220) & (113) & (222) & (004) \\
\colrule
$\Gamma_3 (\psi_{1})$ & 0 & 0 & 0.99 & 1 & 0 & 0 \\
$\Gamma_5 (\psi_{2,3})$ & 1 & 0 & 0.68 & 0.37 & 0 & 0 \\
$\Gamma_7 (\psi_{4,5,6})$ & 1 & 0.74 & 0.34 & 0.37 & 0 & 0 \\
$\Gamma_9 (\psi_{7,9,11})$ & 1 & 0.55 & 0.26 & 0.45 & 0.21 & 0.11 \\
$\Gamma_9 (\psi_{8,10,12})$ & 0.17 & 0.38 & 0.18 & 1 & 0.58 & 0.31 \\
$\Gamma_9$ (L.C.) & 1 & 0.22 & 0.10 & 0.37 & 0.40 & 0.21 \\
\hline
Experiment & 1$\pm$0.2 & 0 & 0.6$\pm$0.3 & 0.4$\pm$0.3 & 0 & 0 \\
\botrule
\end{tabular}
\caption{Powder averaged magnetic Bragg intensities for each of the irreducible representations allowed for Yb$^{3+}$ on the $16d$ site of the $Fd\bar{3}m$ pyrochlore lattice with a propagation vector of $\mathbf{k}=(000)$. Despite having multiple basis vectors, $\Gamma_5$ and $\Gamma_7$ have only one entry because the powder diffraction patterns are identical for each of their basis vectors. The six basis vectors of $\Gamma_9$ likewise produce only two unique powder patterns. We also consider an optimized linear combination of $\Gamma_9$. The final row gives the experimentally observed magnetic intensities. In all cases, the intensities have been normalized relative to the most intense reflection in that pattern.} 
\label{Refinement}
\end{table}

Magnetic Bragg peaks were only found on positions allowed for nuclear scattering by the pyrochlore lattice. Thus, the magnetic reflections in Yb$_2$Ge$_2$O$_7$ can be indexed with a propagation vector of $\mathbf{k}=(000)$. The possible magnetic structures for Yb$^{3+}$ on the $16d$ site of the \emph{Fd$\bar{3}$m} pyrochlore lattice, with propagation vector $\mathbf{k}=(000)$, can be described by four possible irreducible magnetic representations: $\Gamma_{Mag} = \Gamma_3^1 + \Gamma_5^2 + \Gamma_7^3 + \Gamma_9^6$ \cite{Kovalev,0953-8984-18-3-L02}. These irreducible representations can be expressed in terms of their basis vectors ($\psi_1, \psi_2, ...~\psi_{12}$). The $\Gamma_3$ ($\psi_1$) structure is the so-called all-in, all-out state, a non-coplanar antiferromagnetic structure in which the moments are oriented along their local $<$111$>$ axes. The $\Gamma_3$ ($\psi_1$) structure was first experimentally realized in FeF$_3$ \cite{FeF3} and has subsequently been found in various other systems \cite{PhysRevLett.108.247205,PhysRevB.92.144423,doi:10.1143/JPSJ.81.034709}. The $\Gamma_5$ manifold has two basis vectors, $\psi_2$ and $\psi_3$, and has been observed in the $XY$ antiferromagnets Er$_2$Ti$_2$O$_7$ and Er$_2$Ge$_2$O$_7$ \cite{ETO8,PhysRevB.92.140407}. Linear combinations of the $\Gamma_7$ manifold, composed of $\psi_4$, $\psi_5$, and $\psi_6$ are often referred to as the Palmer-Chalker ground state \cite{PhysRevB.62.488}. The Palmer-Chalker ground state is found in Gd$_2$Sn$_2$O$_7$, which is a realization of a Heisenberg pyrochlore antiferromagnet with dipolar interactions \cite{0953-8984-18-3-L02}. There are six basis vectors that make up the $\Gamma_9$ manifold. Linear combinations of these six basis vectors can give non-collinear ferromagnetic structures related to the spin ice-state, such as the splayed ferromagnetic state found in Yb$_2$Sn$_2$O$_7$ \cite{YSO1}.

The simulated relative intensities for each of these representations are listed in Table~\ref{Refinement}. In the case of $\Gamma_5$ and $\Gamma_7$, the powder diffraction patterns for their specific basis vectors are, in general, identical. Thus, we do not distinguish between $\psi_2$ and $\psi_3$, nor do we distinguish between $\psi_4$, $\psi_5$, and $\psi_6$. The six basis vectors that make up $\Gamma_9$ produce two distinct powder diffraction patterns, as indicated in the table. Finally, a linear combination of the $\Gamma_9$ basis vectors can also be considered. While there is poor agreement between $\Gamma_9$ and the experimental data, the linear combination presented in Table~\ref{Refinement} is the one that most closely fits the experimental data: $\Gamma_{Mag}~=~0.038(\psi_{7,9,11}) + 0.021(\psi_{8,10,12})$. Inspection of this table reveals excellent agreement between the experimental results and the $\Gamma_5$ manifold. 

\begin{figure}[tbp]
\linespread{1}
\par
\includegraphics[width=3.3in]{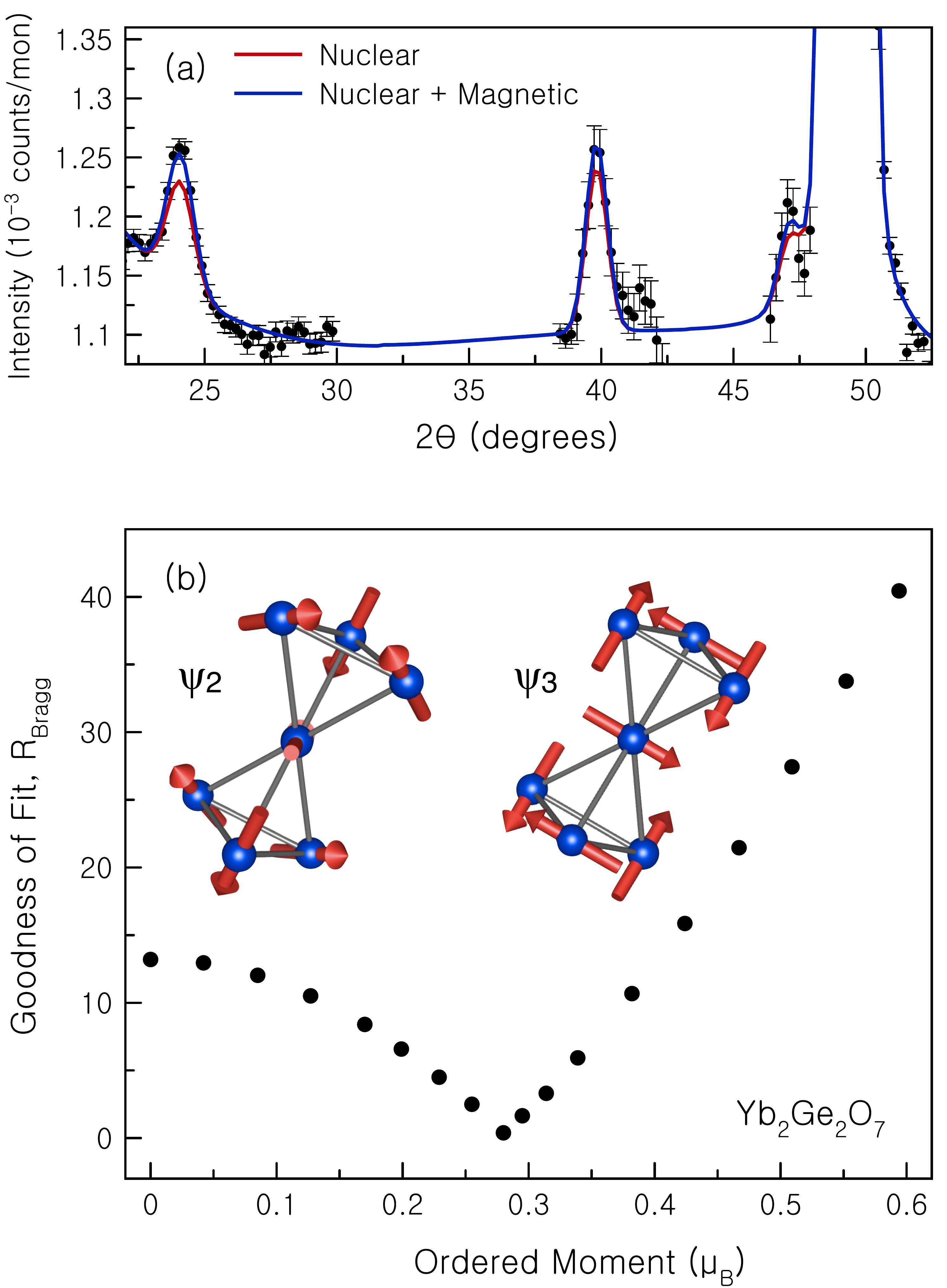}
\par
\caption{(a) Rietveld refinement of Yb$_2$Ge$_2$O$_7$ at 50~mK. All structural and scaling parameters were fixed from a refinement of the 900~mK dataset, as indicated by the red curve. Thus, in the magnetic refinement, indicated by the blue curve, the only adjustable parameter is the moment size. (b) Goodness of fit of the magnetic structure refinement for Yb$_2$Ge$_2$O$_7$, as measured by the minimization of $R_{\text{Bragg}}$, as a function of the magnitude of the ordered moment. $R_{\text{Bragg}}$ is the sum of the weighted residuals for only the magnetic Bragg reflections. The best agreement between is given by an ordered moment of 0.3(1)~$\mu_B$. The inset shows the spin alignments given by the $\psi_2$ and $\psi_3$ states.}
\label{Moment_Dep}
\end{figure}

Rietveld refinement of all measured Bragg reflections, as summarized in Table 2, was used to determine the size of the ordered moment in Yb$_2$Ge$_2$O$_7$. All structural parameters for the pyrochlore $Fd\bar{3}m$ lattice and the scaling factor were determined from a refinement of the 900~mK data set, which is well above $T_N$. The 50~mK data set was then refined with a $\Gamma_5$ magnetic structure where only the magnitude of the ordered moment was allowed to freely vary. The $\psi_2$ and $\psi_3$ basis vectors which comprise $\Gamma_5$ generate identical powder neutron diffraction patterns and identical magnetic moment sizes, and thus we do not distinguish between the two within our Rietveld refinement. The resultant Rietveld refinement for Yb$_2$Ge$_2$O$_7$ at 50~mK is shown in Figure~\ref{Moment_Dep}(a). The best agreement with the measured data, as indicated by a minimization of $R_{\text{Bragg}}$, was obtained for an ordered moment of 0.3(1)~$\mu_{\text{B}}$ (Figure~\ref{Moment_Dep})(b).

\begin{figure}[tbp]
\linespread{1}
\par
\includegraphics[width=3.3in]{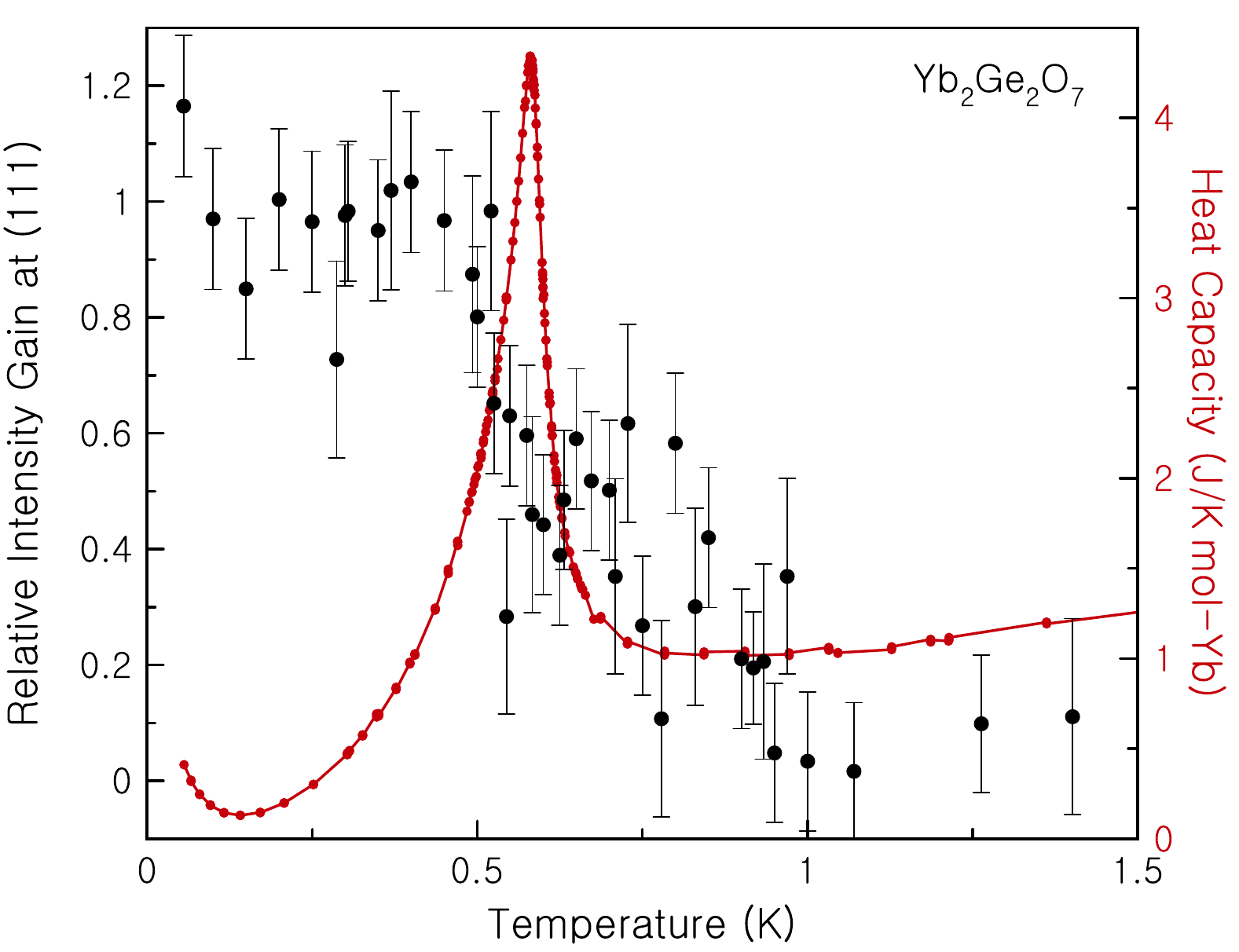}
\par
\caption{The relative intensity of the (111) magnetic Bragg reflection as a function of temperature. The intensity is given relative to the average intensity between 1~K and 5~K, which is set to zero. The order parameter correlates well with the heat capacity anomaly, shown in red on the right hand axis. The upturn below 100~mK in the heat capacity corresponds to a nuclear Schottky anomaly.}
\label{YGO_OP}
\end{figure}

In order to obtain a measurement of the order parameter in Yb$_2$Ge$_2$O$_7$, we tracked the (111) Bragg peak, which is the largest magnetic reflection. Figure \ref{YGO_OP} shows the intensity of (111) as a function of temperature, where the zero has been set by the average intensity between 1~K and 5~K, well above the N\'eel temperature. The order parameter in Yb$_2$Ge$_2$O$_7$ correlates directly with the sharp anomaly in the heat capacity, which is peaked at $T_N$~=~0.57~K. While the order parameter in Yb$_2$Ge$_2$O$_7$ appears quite conventional, this is not generically true in the ytterbium pyrochlores. The order parameter in Yb$_2$Ti$_2$O$_7$, which plateaus below $T_c=240$~mK, continually decreases well above $T_c$, to at least 700~mK \cite{Jonathan_YTO}.

\subsection{IV. Comparison to Relevant XY Pyrochlore Magnets}

The antiferromagnetic $\Gamma_5$ structure we have determined for Yb$_2$Ge$_2$O$_7$ below $T_N=0.57$~K belongs to the same ground state manifold as Er$_2$Ti$_2$O$_7$ below its $T_N=1.2$~K transition. However, Er$_2$Ti$_2$O$_7$ was identified as uniquely displaying the $\psi_2$ state, with a rather large ordered moment of $\mu_{\text{ord}}=$~3.01~$\mu_B$ \cite{ETO8}. For our powder sample of Yb$_2$Ge$_2$O$_7$, we cannot distinguish $\psi_2$ from $\psi_3$ within $\Gamma_5$ (Inset of Figure~\ref{Moment_Dep}(b)), and the ordered moment within this structure at low temperatures is small, $\mu_{\text{ord}}=$~0.3(1)~$\mu_B$. We note that a large ordered moment of 1.06(7)~$\mu_B$ has recently been reported for the antiferromagnetic ground state of Yb$_2$Ge$_2$O$_7$ \cite{PhysRevB.92.140407}, but this estimate arose from measurements on a much smaller volume of sample, and no net magnetic scattering (\emph{ie.} difference between high and low temperature) is shown for the strongest magnetic Bragg peak, (111). In any case, this large ordered moment estimate for Yb$_2$Ge$_2$O$_7$ is inconsistent with our results (Figure~\ref{Moment_Dep}(b)).

Order-by-quantum disorder has been proposed as the mechanism for the selection of the $\psi_2$ ground state for Er$_2$Ti$_2$O$_7$ \cite{ETO6,ETO7,ETO8,ETO9,ETO14,ETO15}, based on understanding the microscopic spin Hamiltonian derived from spin wave measurements \cite{ETO6}. This mechanism predicts a gap in the spin wave spectrum due to breaking the continuous symmetry which exists between the $\psi_2$ and $\psi_3$ ground states within $\Gamma_5$. Indeed, this spin wave gap has been measured in Er$_2$Ti$_2$O$_7$ \cite{ETO10}. However, an alternative mechanism for ground state selection based on virtual transitions to excited crystal field levels has also been proposed for Er$_2$Ti$_2$O$_7$ \cite{ETO12,ETO13,ETO15}. This alternative mechanism relies on the presence of low energy crystal field levels, as the probability for such virtual transitions go as the inverse square of the energy required for the transitions out of the ground state. This is a plausible scenario for Er$_2$Ti$_2$O$_7$, as the lowest lying crystal field levels in Er$_2$Ti$_2$O$_7$ are at 6.3~meV and 7.3~meV \cite{ETO8}. However, it is not a plausible scenario for Yb$_2$Ge$_2$O$_7$, as we have just determined that the lowest crystal field transition occurs at 80.7~meV, more than an order of magnitude higher in energy than was the case for Er$_2$Ti$_2$O$_7$. In this regard, Yb$_2$Ge$_2$O$_7$ is a stronger candidate for exhibiting an ordered state selected by a thermal or quantum order-by-disorder mechanism.

We emphasize that it does not follow that the $\Gamma_5$ antiferromagnetic ground state we observe in Yb$_2$Ge$_2$O$_7$ arises from an order-by-disorder mechanism. It has  been shown that a $\Gamma_5$ state in Yb$_2$Ge$_2$O$_7$ could be predicted purely on the basis of the phase diagram for Yb$_2$Ti$_2$O$_7$ obtained from its anisotropic spin exchange Hamiltonian \cite{2015arXiv150505499J}. Indeed, as $\Gamma_5$ is constituted by both $\psi_2$ and $\psi_3$, it is not clear that a selection is even being made, which would necessitate an order-by-disorder scenario. Dun \emph{et al.} claim that fits to the heat capacity below $T_N=$~0.57~K are consistent with the presence of an emerging spin wave gap of 24~$\mu$eV, but no such gap has been directly measured. Nonetheless, the absence of low lying crystal field excitations in Yb$_2$Ge$_2$O$_7$ imply that there could be significant differences between the antiferromagnetic ground states in Yb$_2$Ge$_2$O$_7$ and Er$_2$Ti$_2$O$_7$, despite their similarities.


\section{Conclusions}

To conclude, we have synthesized relatively large volumes of the pyrochlore magnet Yb$_2$Ge$_2$O$_7$ using high pressure synthesis techniques. This has enabled studies of both the crystal field excitations of Yb$_2$Ge$_2$O$_7$ using inelastic neutron scattering, and the low temperature ground state of this system, using magnetic neutron diffraction and $\mu$SR techniques. Our inelastic neutron scattering measurements allow us to determine the eigenvalues and eigenfunctions associated with the splitting of the (2$J$+1) manifold of states appropriate to Yb$^{3+}$ in the Yb$_2$Ge$_2$O$_7$ environment. We find an XY nature to the single ion ground state wavefunction, as expressed in $g_{\perp}/g_z =$~1.7(2), and a large 80.7~meV gap to the first excited state. The ground state doublet is primarily comprised of $m_J=\pm \sfrac{1}{2}$, supporting a picture of $S_{\text{eff}}=\sfrac{1}{2}$ Yb$^{3+}$ moments. 

$\mu$SR measurements show quasi-static magnetic order on the muon time scale to set in below $T_N=$~0.57~K. Our elastic neutron scattering measurements show the ground state to be a $\Gamma_5$, $\mathbf{k}=(000)$, antiferromagnetic state, with a relatively small ordered moment of 0.3(1)~$\mu_B$ at low temperatures. We hope that this characterization of the single ion and ground state properties of Yb$_2$Ge$_2$O$_7$ motivates a full understanding of the structure and dynamics of this exotic pyrochlore magnet, and helps guide a thorough understanding of its fascinating phase behavior.

\begin{acknowledgments}

\section{Acknowledgments}
We acknowledge useful conversations with M.J.P. Gingras and J. Rau. We appreciate the hospitality of the TRIUMF Centre for Molecular and Materials Science and thank B.S. Hitti, G.D. Morris and D.J. Arseneau for assistance with the $\mu$SR measurements. A.M.H.
acknowledges support from the Vanier Canada Graduate Scholarship Program and thanks the National Institute for Materials Science (NIMS) for their hospitality and support through the NIMS Internship Program. This work was supported by the Natural Sciences and Engineering Research Council of Canada and the Canada Foundation for Innovation. The research at HFIR and SNS, ORNL, was sponsored by the Scientific User Facilities Division, Office of Basic Energy Sciences, US Department of Energy. Work at the University of Edinburgh was supported by EPSRC and the Royal Society. R.S.F. acknowledges support from CNPq (Grant No. 400278/2012-0).  

\end{acknowledgments}

\bibliography{YGO_Ref_2}

\end{document}